\documentclass[superscriptaddress,preprintnumbers,amsmath,amssymb,twocolumn,floats]{revtex4-1}
\usepackage{txfonts}
\usepackage{amssymb}
\usepackage{amsmath}
\usepackage{graphicx}
\usepackage{sidecap}
\usepackage{CJK}
\usepackage{url}
\usepackage{color}
\usepackage{gensymb}
\usepackage{setspace}
\usepackage[multi-part-units = single]{siunitx}
\makeatletter

\newcommand{\Rmnum}[1]{\expandafter\@slowromancap\romannumeral #1@}
\makeatother
\begin{document}
\title{Evidence of topological nodal lines and surface states\\ in the centrosymmetric superconductor SnTaS$_2$}

\author{Wenqing Chen}
%%\email{wqchen@mail.sim.ac.cn}
\affiliation{State Key Laboratory of Functional Materials for Informatics and Center for Excellence in Superconducting Electronics, Shanghai Institute of Microsystem and Information Technology, Chinese Academy of Sciences, Shanghai 200050, China}
\affiliation{College of Materials Science and Opto-Electronic Technology, University of Chinese Academy of Sciences, Beijing 100049, China}

\author{Lulu Liu}
%%\email{dg1922034@smail.nju.edu.cn}
\affiliation{National Laboratory of Solid State Microstructures, School of Physics, Nanjing University, Nanjing 210093, China}

\author{Wentao Yang}
%%\email{wtyang18@fudan.edu.cn}
\affiliation{State Key Laboratory of Surface Physics, Department of Physics, and Advanced Materials Laboratory, Fudan University, Shanghai 200438, China}

\author{Dong Chen}
%%\email{dchen@qdu.edu.cn}
\affiliation{College of Physics, Qingdao University, Qingdao 266071, China}

\author{Zhengtai Liu}
%%\email{ztliu@mail.sim.ac.cn}
\affiliation{State Key Laboratory of Functional Materials for Informatics and Center for Excellence in Superconducting Electronics, Shanghai Institute of Microsystem and Information Technology, Chinese Academy of Sciences, Shanghai 200050, China}

\author{Yaobo Huang}
%%\email{huangyaobo@zjlab.org.cn}
\affiliation{Shanghai Advanced Research Institute, Chinese Academy of Sciences, Shanghai 201204, China}

\author{Tong Zhang}
%%\email{tzhang18@fudan.edu.cn}
\affiliation{State Key Laboratory of Surface Physics, Department of Physics, and Advanced Materials Laboratory, Fudan University, Shanghai 200438, China}
\affiliation{Collaborative Innovation Center of Advanced Microstructures, Nanjing University, Nanjing 210093, China}

\author{Haijun Zhang}
\email{zhanghj@nju.edu.cn}
\affiliation{National Laboratory of Solid State Microstructures, School of Physics, Nanjing University, Nanjing 210093, China}
\affiliation{Collaborative Innovation Center of Advanced Microstructures, Nanjing University, Nanjing 210093, China}

\author{Zhonghao Liu}
\email{lzh17@mail.sim.ac.cn}
\affiliation{State Key Laboratory of Functional Materials for Informatics and Center for Excellence in Superconducting Electronics, Shanghai Institute of Microsystem and Information Technology, Chinese Academy of Sciences, Shanghai 200050, China}
\affiliation{College of Materials Science and Opto-Electronic Technology, University of Chinese Academy of Sciences, Beijing 100049, China}

\author{D. W. Shen}
\email{dwshen@mail.sim.ac.cn}
\affiliation{State Key Laboratory of Functional Materials for Informatics and Center for Excellence in Superconducting Electronics, Shanghai Institute of Microsystem and Information Technology, Chinese Academy of Sciences, Shanghai 200050, China}
\affiliation{College of Materials Science and Opto-Electronic Technology, University of Chinese Academy of Sciences, Beijing 100049, China}

\begin{abstract}
The discovery of signatures of topological superconductivity in superconducting bulk materials with topological surface states has attracted intensive research interests recently. Utilizing angle-resolved photoemission spectroscopy and first-principles calculations, here, we demonstrate the existence of topological nodal-line states and drumheadlike surface states in centrosymmetric superconductor SnTaS$_2$, which is a type-\Rmnum{2} superconductor with a critical transition temperature of about 3 K. The valence bands from Ta 5$d$ orbitals and the conduction bands from Sn 5$p$ orbitals cross each other, forming two nodal lines in the vicinity of the Fermi energy without the inclusion of spin-orbit coupling (SOC), protected by the spatial-inversion symmetry and time-reversal symmetry. The nodal lines are gapped out by SOC. The drumheadlike surface states, the typical characteristics in nodal-line semimetals, are quite visible near the Fermi level. Our findings indicate that SnTaS$_2$ offers a promising platform for exploring the exotic properties of the topological nodal-line fermions and helps in the study of topological superconductivity.
\end{abstract}
\maketitle

\section{Introduction}
The long-sought topological superconductor (TSC) with Majorana fermions extends scientific territory and plays an important role in strategies for topological quantum computing, though it still remains elusive in real materials. A typical TSC can be found in a $p+ip$ system, where the core of the superconducting vortex contains a localized quasiparticle with exactly zero energy \cite{Qi2011}. However, the properties of several experimental candidates for $p$-wave superconductivity remain unclear \cite{Mackenzie2003}. Besides the $p$-wave pairing, the TSC could also be realized in the system with an $s$-wave pairing plus the spin-orbit coupling (SOC) and spin polarization, which is equivalent to the $p+ip$ system with broken spin-rotation symmetry \cite{Fu2008,Liu2011}. Due to the proximity effect \cite{Fu2008,Huang2019}, an effective $p$-wave pairing can be realized on the interface of a heterostructure between a strong topological insulator (TI) and an $s$-wave superconductor (e.g., Bi$_2$Se$_3$/NbSe$_2$, Bi$_2$Te$_3$/NbSe$_2$, and Bi$_2$Se$_3$/Pb) \cite{Mourik2012,Wang2012,Xu2015,Sun2016,Qu2012} or a full-gapped bulk superconductor with topologically protected gapless surface (or edge) states (e.g., $\beta$-PdBi$_2$, FeTe$_{1-x}$Se$_x$, Cu$_x$Bi$_2$Se$_3$, Li$_{0.84}$Fe$_{0.16}$OHFeSe, and 2$M$-WS$_2$) \cite{Sakano2015,Hao2019,Sasaki2011,Tao2018,Wang2018,Zhang2018,Yin2015,Liu2018,WS_Huang,WS_QKXue}. Based on the latter scenario, type-\Rmnum{2} superconductors with nontrivial surface states are promising candidates for TSCs. The surface states and superconducting gaps can be experimentally determined by angle-resolved photoemission spectroscopy (ARPES) and quasiparticle scattering interference imaging combined with scanning tunneling spectra. The scanning tunneling spectra can further be used to search for Majorana zero modes in the vortex core of these materials.

In addition, noncentrosymmetric superconductors with strong SOC have been proposed as a kind of TSC, because of the breaking of spin degeneracy by asymmetric SOC \cite{Ali2014}. It has been reported that noncentrosymmetric PbTaSe$_2$ hosting topological nodal lines and nontrivial surfaces states around the Fermi energy ($E\rm_F$) is a candidate for TSC without the aid of doping or artificial heterostructures \cite{Bian2016,Chang2016,Wang2016,Guan2016,Xu2019,Ikeda2020,Benia2016}. The superconductivity on its surface states induced by the proximity effect can be stronger than that in TI/$s$-wave superconductor heterostructures \cite{Chang2016}. Combining its helical spin-polarized surface state \cite{Chang2016} with superconductivity, PbTaSe$_2$ therefore would be a TSC even without asymmetric SOC. Here, we find the centrosymmetric superconductor SnTaS$_2$, which is isoelectronic to PbTaSe$_2$ and also has a crystal structure similar to that of layered PbTaSe$_2$. This centrosymmetric crystal structure is protected by the spatial-inversion symmetry and the time-reversal symmetry ($T$$\cdot$$P$ symmetry) \cite{Chiu2014,Yang2018}. The previous calculations suggest that SnTaS$_2$ has nodal-line structures with drumheadlike surface states around $E\rm_F$ \cite{Chen2019,Dijkstra1989,Jin2019}, making it a good candidate to further detect the TSCs and Majorana fermions. However, so far, direct ARPES measurements of the low-energy electronic states are still absent.

\begin{figure}[t!]
\centering
\includegraphics[width=0.47\textwidth]{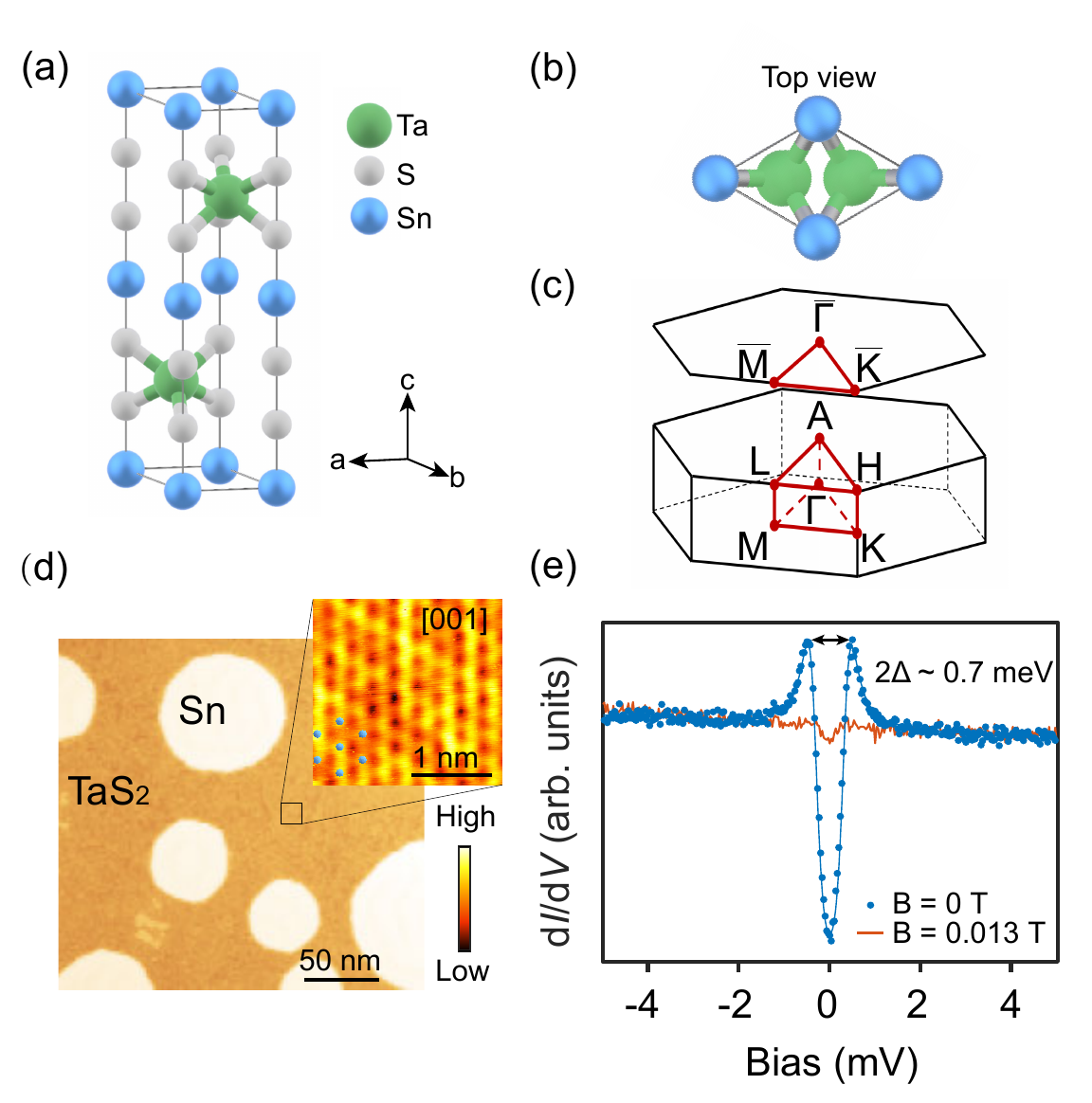}
\caption{
(a) Crystal structure of SnTaS$_2$.
(b) The top view of the crystal structure shows centrosymmetric structure.
(c) Three-dimensional bulk and projected 2D BZs.
(d) STM topography image of the cleaved SnTaS$_2$ surface at $T$ = 4 K. Set point condition: $V$ = 3 V, $I$ = 10 pA. Inset: Atomic-resolved STM topography taken on TaS$_2$ surface shows hexagonal atomic structure. Set point condition: $V$ = -0.6 V, $I$ = 10 pA.
(e) Scanning tunneling spectra ($dI/dV$) taken on the TaS$_2$ surface at $T$ = 0.4 K under a magnetic field of 0 (blue) and 0.013 T (red) perpendicular to the sample surface. The superconducting gap of about 0.35 meV is estimated by fitting the blue points with the Dynes function.
}
\label{1}
\end{figure}

In this paper, we demonstrate the existence of topological nodal-line states and nontrivial surface states in a type-\Rmnum{2} BCS superconductor SnTaS$_2$ by employing ARPES and scanning tunneling microscopy/spectroscopy (STM/S) in combination with first-principles calculations. STS measurements indicate that the superconducting gap of about 0.35 meV at $T$ = 0.4 K can be suppressed by an applied magnetic field of 0.013 T perpendicular to the sample surface, which is consistent with the transport measurements \cite{Chen2019}. The calculations show that the nodal lines are formed by the crossings of the Ta 5$d$ orbitals and the Sn 5$p$ orbitals in the vicinity of $E\rm_F$ without the inclusion of SOC. If SOC is considered, nodal lines are gapped out. The surface bands connecting to each nodal line indicate the topological nontrivial states. ARPES observations of the features are in good agreement with the band calculations. The observed surface state can be clearly distinguished from the bulk states near $E\rm_F$. These results uncover the topological electronic states in superconducting SnTaS$_2$, offering a new platform for further studies into Dirac nodal-line physics as well as TSCs.

\begin{figure*}[ht!]
\centering
\includegraphics[width=0.95\textwidth]{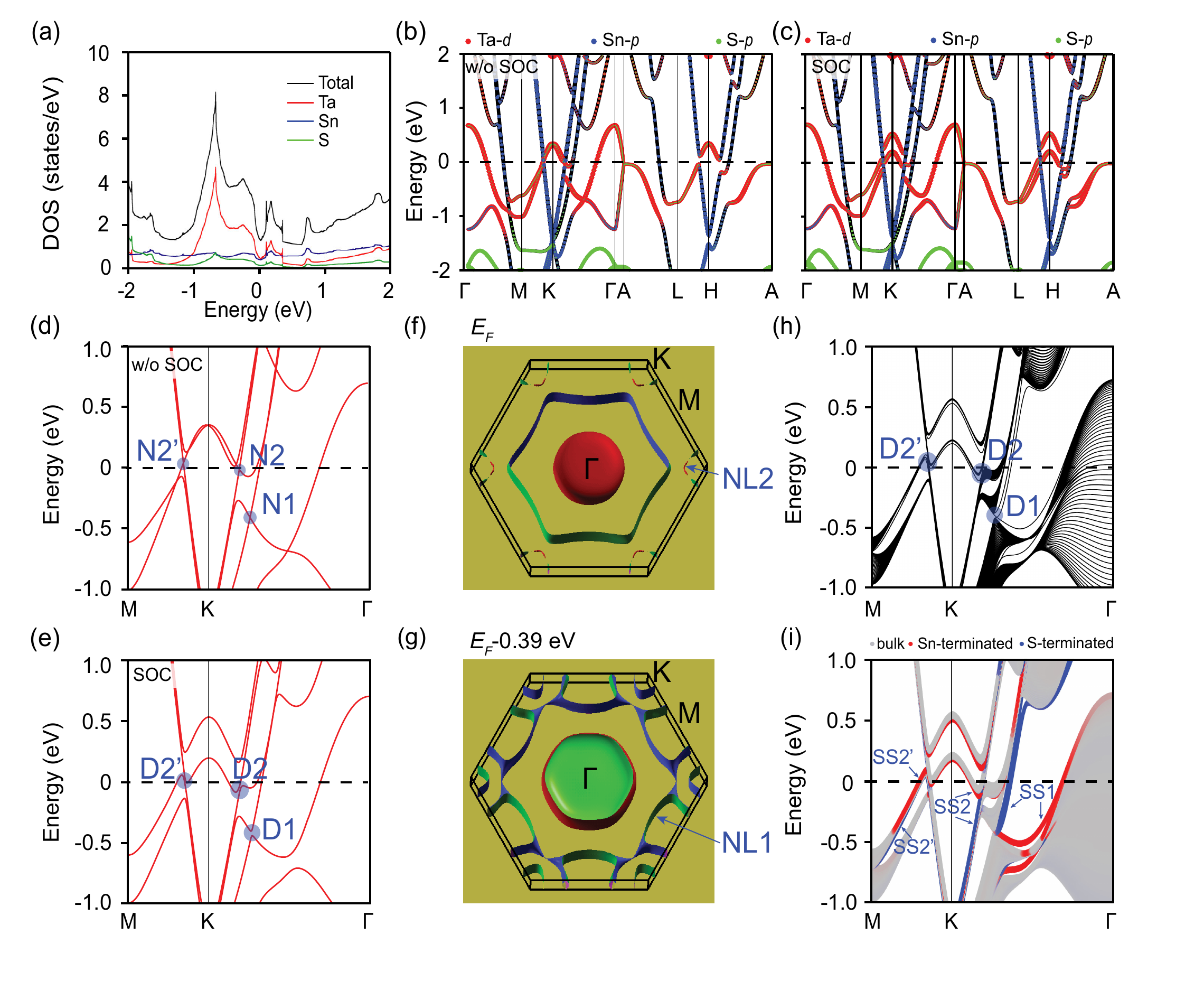}
\caption{
(a) Total and orbital-projected DOS within $\pm$2 eV at $E\rm_F$. The different colors indicate the different orbitals.
(b) and (c) The calculated bulk bands with their orbital characters along the high-symmetry directions without and with SOC, respectively.
(d),(e) Enlarged views of bulk band structure without and with SOC on the $M$-$K$-$\Gamma$ plane around $E\rm_F$, respectively, show nodal-line structure. The band crossings in (d) are indicated by N1, N2, and N2$^\prime$. The gapped nodes in (e) are indicated by D1, D2, and D2$^\prime$.
(f) and (g) The calculated constant-energy surfaces without SOC at $E\rm_F$ and $E\rm_F$ - 0.39 eV, respectively, show nodal lines (NL1 and NL2) as indicated by the arrows.
(h) and (i) Band structures with SOC for the 30-unit-thick layer. The bands with separation and the gray shaded regions present the bulk bands, and the others are surface bands. Projected band structures for the Sn-terminated (red) and S-terminated (blue) (001) surfaces of SnTaS$_2$ are shown in (i). The surface states connecting the nodes are indicated by the arrows.
}
\label{2}
\end{figure*}

\begin{figure*}[ht!]
\includegraphics[width=0.9\textwidth]{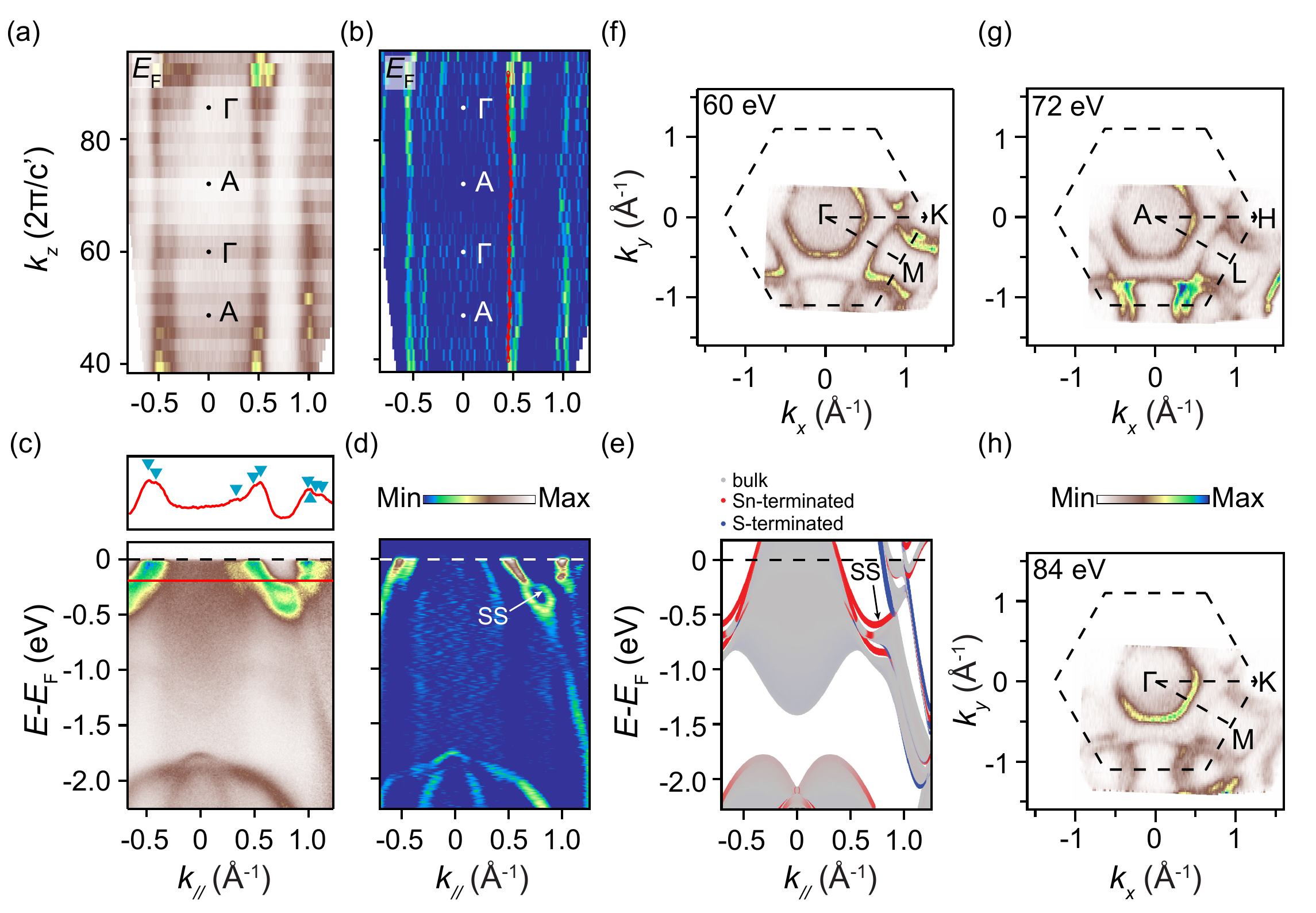}
\caption{
(a) Integrated intensity plots on the \emph{k$_{z}$}-\emph{k$_\parallel$} plane at $E_F\pm$10 meV.
(b) Corresponding second-derivative plots of (a). The red line indicates a band with weak $k_z$ dispersion.
(c) Bottom: Photoemission intensity plots taken along the $A$-$H$ direction with 48-eV photons ($k_z\sim$ $\pi$). Top: MDC is taken along the red solid line, and the peaks are indicated by the turquoise arrows.
(d) Corresponding second-derivative plots of (c). The observed surface state is indicated by the arrow.
(e) Projected band structures for the two terminated (001) surfaces along $\bar{\Gamma}-\bar{K}$.
(f)-(h) Integrated intensity plots at $E_F\pm$10 meV taken with 60-eV ($k_z\sim$ 0), 72-eV ($k_z\sim$ $\pi$), and 84-eV ($k_z\sim$ 0) photons, respectively.
}
\label{3}
\end{figure*}

\begin{figure*}[ht!]
\centering
\includegraphics[width=0.95\textwidth]{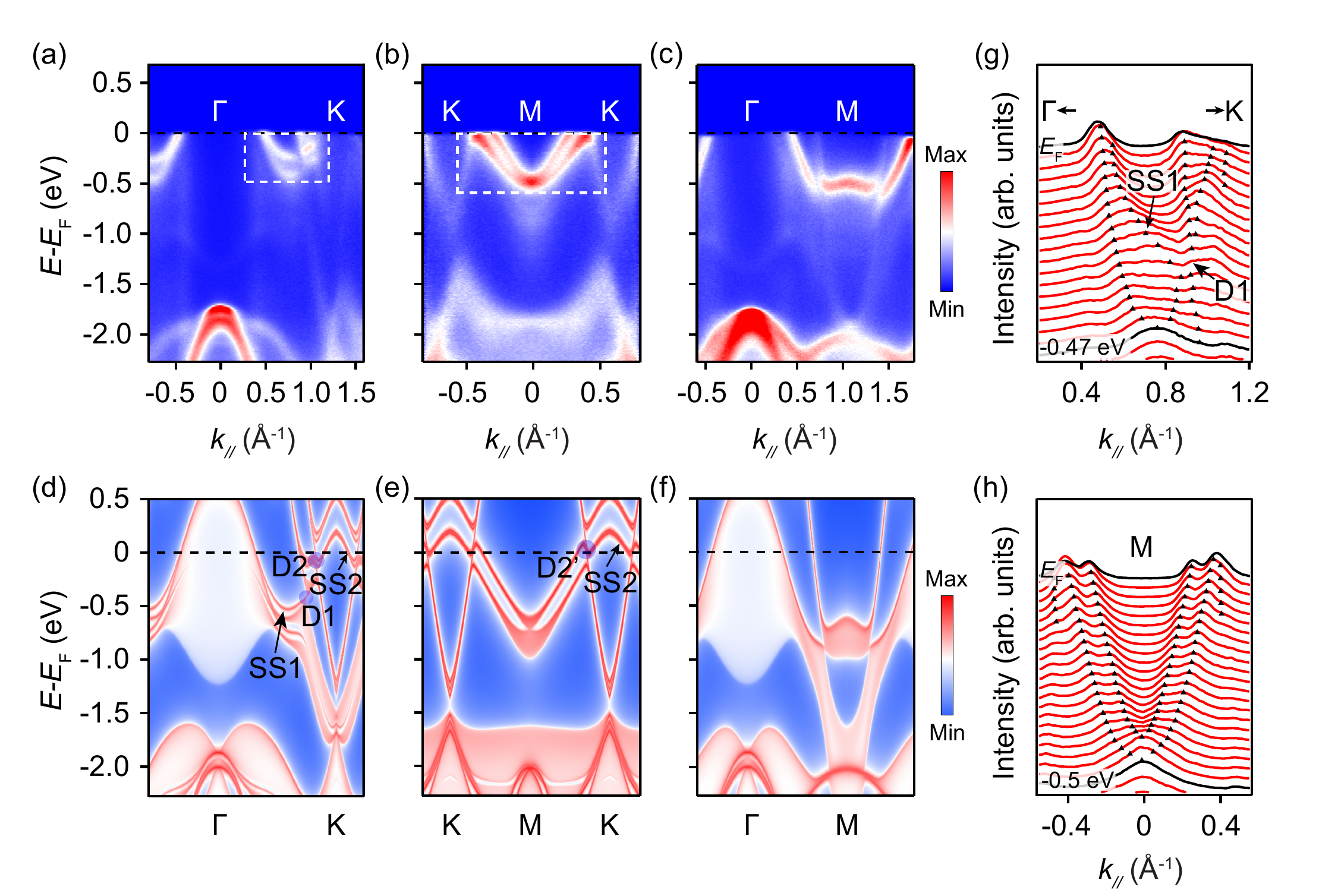}
\caption{
(a)-(c) Photoemission intensity plots along the $\Gamma$-$K$, $K$-$M$, and $\Gamma$-$M$ directions, respectively.
(d)-(f) DFT-projected bulk bands and surface bands of the Sn-terminated (001) surface along the $\Gamma$-$K$, $K$-$M$, and $\Gamma$-$M$ directions, respectively, corresponding to (a)-(c). The surface states and the Dirac nodes are indicated by the black arrows and the transparent dots, respectively.
(g) MDCs around the nodal-line feature near the $K$ point along $\Gamma$-$K$, as indicated by the dashed rectangle in (a).
(h) MDCs around the nodal-line feature near the $K$ point along $K$-$M$, as indicated by the dashed rectangle in (b). The dashed lines are guides to the bands.
}
\label{4}
\end{figure*}

\section{Methods}
The single crystals were grown by the chemical vapor transport method as described elsewhere \cite{Chen2019}. STM experiments were carried out with a Unisoku low-temperature STM system. SnTaS$_2$ single crystals were cleaved $in$ $situ$ at $T$ = 77 K under ultrahigh vacuum and transferred immediately to a STM head under a vacuum of 2$\times$10$^{-10}$ Torr. STS measurements were done using a standard lock-in technique with a bias modulation of 0.1 mV at a frequency of 741 Hz. ARPES measurements were performed at the Dreamline and 03U beamlines of Shanghai Synchrotron Radiation Facilities (SSRF). Samples smaller than $1\times1$ mm$^2$ were cleaved $in$ $situ$, exposing flat, mirrorlike (001) surfaces. During measurements, the temperature was kept at $T$ = 20 K, and the pressure was maintained at less than $4\times10^{-11}$ Torr. The first-principles calculations based on the density functional theory (DFT) were performed by using the Vienna Ab initio Simulation Package (VASP) \cite{Kertesz2003,Hafner2007}, with the generalized gradient approximation of the Perdew-Burke-Ernzerhof type as the exchange-correlation functional \cite{Yanai2004,Jain2011}. The plane-wave cutoff energy was set to be 420 eV in all first-principles calculations, and the first Brillouin zone (BZ) was sampled using a $13\times13\times5$ Monkhorst-Pack grid. The experimental lattice constants $a$ = $b$ = 3.309 {\AA} and $c$ = 17.450  {\AA} \cite{Chen2019} were used in the calculations.

\section{Results and discussions}
The crystal structure of SnTaS$_2$ is illustrated in Figs. \ref{1}(a) and \ref{1}(b), and crystallizes in a layered hexagonal structure with a space group of $P$6$_3$/$mmc$ (No. 194).  Isoelectronic PbTaSe$_2$ [$P\bar{6}m_2$ (No. 187)] has a similar layered hexagonal structure. SnTaS$_2$ is a centrosymmetric crystal with $T$$\cdot$$P$ symmetry, however, noncentrosymmetric PbTaSe$_2$ without $T$$\cdot$$P$ symmetry has a reflection symmetry caused by the Ta atomic plane \cite{Bian2016,Ali2014}. The corresponding three-dimensional (3D) and projected two-dimensional (2D) BZs ($\bar{\Gamma}-\bar{K}-\bar{M}$) are shown in Fig. \ref{1}(c), where the $\Gamma$, $K$, and $M$ points serve as the high-symmetry points of the $k_z$ = 0 plane and the $A$, $H$, and $L$ points serve as the high-symmetry points of the $k_z$ = $\pi$ plane.

To reveal the surface topography and the superconductivity, STM/S measurements were carried out on the (001) surfaces of SnTaS$_2$. The high-resolution STM topography taken on the area of the cleaved surface [Fig. \ref{1}(d)] shows a flat TaS$_2$ plane decorated with islands of Sn. The atomic-resolved STM topography taken on the TaS$_2$ surface [inset of Fig. \ref{1}(d)] shows a clear hexagonal atomic structure with a lattice parameter of $a$ = 3.3 {\AA}. Figure \ref{1}(e) displays the scanning tunneling spectra ($dI/dV$) at $T$ = 0.4 K under a magnetic field of 0 (blue) and 0.013 T (red) perpendicular to the sample surface. To determine the gap size more precisely, we fit the tunneling spectrum with the Dynes function. The fitting curve is obtained by the thermal broadened density of states (DOS), which was calculated from the tunneling current expression \cite{Sacks1991,Sugimoto2015}:
\begin{spacing}{2.0}
 \par
 $I(V) \propto \int_{-\infty}^{\infty}\left|\operatorname{Re} \frac{E-i \Gamma}{\sqrt{(E-i \Gamma)^{2}-\Delta^{2}}}\right|[f(E)-f(E+e V)] d E$
 \par
\end{spacing}
\noindent where $f(E)$ is the Fermi-Dirac distribution function. The superconducting gap of about 0.35 meV at $T$ = 0.4 K is estimated based on the fitting curve [blue curve in Fig. \ref{1}(e)]. By applying a magnetic field of 0.013 T perpendicular to the $ab$ plane, the superconducting gap is suppressed. No vortex state can be found before the gap is closed, although many attempts have been made in various regions of different surfaces and samples. Considering the small upper critical field of $H \parallel c$ [$H_{c2} (0)$ = 203.6 Oe] \cite{Chen2019}, the vortex state is hard to find, possibly due to the low vortex density under such a small applied magnetic field.

We performed orbital-projected DFT calculations to study the low-energy electronic bands with their orbital characters in SnTaS$_2$.  Figure \ref{2}(a) shows total and orbital-projected DOS within $\pm$2 eV at $E\rm_F$. The obvious dip of DOS near $E\rm_F$ suggests a semimetallic character. Figures \ref{2}(b) and \ref{2}(c) show orbital-projected dispersions along the high-symmetry directions without and with SOC, respectively. The electronic structure near $E\rm_F$ is mainly dominated by Ta 5$d$ (red) and Sn 5$p$ (blue) orbitals mixed with less S 3$p$ (green). The holelike Ta 5$d$ bands and the electronlike Sn 5$p$ bands cross each other, forming several band crossings around the $K$ point near $E\rm_F$, as shown in Fig. \ref{2}(b). When SOC is taken into account, band gaps open at the crossing points, as shown in Fig. \ref{2}(c). In order to see the crossing points clearly, we enlarge the view around the $K$ point near $E\rm_F$, as shown in Figs. \ref{2}(d) and \ref{2}(e). Without the inclusion of SOC, there are two crossing points considered as nodes along the $\Gamma$-$K$ direction, namely, N1 and N2. The corresponding crossing point of N2 along the $\Gamma$-$M$ direction is marked with N2$^\prime$. With the inclusion of SOC, the nodes are all gapped out. The sizes of SOC gaps at D1 and D2 (D2$^\prime$) are about 40 and 80 meV [Fig. \ref{2}(e)], respectively, which are comparable with typical nodal-line materials such as the PbTaSe$_2$, CaAgAs, TiB$_2$, ZrSiS, and CaP$_3$ families \cite{Bian2016,Yamakage2016,Liu2018b,ZrB_Lou,Takane2016,Topp2017,Fu2019,Xu2017,Song2020}. To carefully check whether the band crossings form the nodal lines in the momentum space, we calculated constant- energy surfaces without SOC at $E\rm_F$ and $E\rm_F$ - 0.39 eV, respectively. The nodal lines (NL1 and NL2) formed by N1 and N2 (N2$^\prime$) are indicated by the arrows, as shown in Figs. \ref{2}(f) and \ref{2}(g). Because the energy positions of crossing points of NL2 are different surrounding the $K$ point, NL2 shows various thicknesses at different momentum positions on the constant-energy surface.

In addition, the drumheadlike surface state is a typical characteristic in nodal-line semimetals and could bring varieties of unusual intriguing properties. To study the surface states, we performed the slab calculation with a 30-unit-thick layer [Figs. \ref{2}(h) and \ref{2}(i)]. The bands with separation and the gray shaded regions present the bulk bands, and the others are surface bands in Figs. \ref{2}(h) and \ref{2}(i). Figure \ref{2}(i) shows the projected spectra for the Sn-terminated (red) and S-terminated (blue) (001) surfaces of SnTaS$_2$. The clear surface states connecting the nodes are indicated by the arrows for the two terminated (001) surfaces. Due to the absence of chiral symmetry in this material, the drumheadlike surface state is sensitive to surface conditions and is not flat. However, one can always find the surface bands connecting to the nodes in the calculation, indicative of the nontrivial topological nature of the nodal lines.

To directly detect the topological features near $E\rm_F$, we performed ARPES measurements on (001) surfaces of SnTaS$_2$. Since the nodal lines are located on the $k_z$ = 0 plane according to the calculation, we carried out the photon-energy-dependent ARPES measurement to investigate the detailed $k_z$ dispersions. With an empirical value of the inner potential of $\sim$11 eV and $c$$^\prime$ = $c$/2 = 8.725 {\AA} (due to the bilayer in the conventional unit cell), we found that $hv$ = 60 and 84 eV are close to the $\Gamma$ point and 48 and 72 eV are close to the $A$ point, according to the free-electron final-state model \cite{Liu2012}. Although the bands near $E\rm_F$ show weak $k_z$ dispersions in Figs. \ref{3}(a) and \ref{3}(b), a periodic modulation along the $k_z$ direction can still be found, as shown by the dispersions marked by the red line in Fig. \ref{3}(b). Figures \ref{3}(c) and \ref{3}(d) show energy-momentum plots and corresponding second derivative plots along the $A$-$H$ line (48 eV), respectively. The marks on the momentum distribution curve (MDC) taken at a binding energy of about 0.2 eV refer to the bands crossing $E\rm_F$. The whole band structure looks more like the slab calculation with Sn-terminated (001) surface states of Fig. \ref{2}(i), rather than the bulk states of Fig. \ref{2}(c). Especially, the bands around the BZ center with weak intensities seem to be stuck together, which is different from the clean bulk-calculation bands with strong dispersions along the $\Gamma-A$ direction in Fig. \ref{2}(c). This discrepancy is mainly caused by the $k_z$ integration and matrix element effects in the ARPES spectra, which reflect the electronic states integrated over a certain $k_z$ region of the bulk BZ and the states at $k_z$ = 0 and $\pi$ contributed greatly \cite{Kumigashira1998}. As evidenced in Figs. \ref{3}(f)-\ref{3}(h), the Fermi surfaces of the $k_z$ $\sim$ 0 and $\pi$ planes are almost the same. The Fermi surfaces are similar to those of PbTaSe$_2$, which can be more precisely figured out with the help of calculations and quantum oscillations \cite{Xu2019}. In recent studies on the topological semimetals  \cite{Bian2016,Takane2016,Topp2017,Liu2018b,ZrB_Lou}, it is noticeable that the states from $k_z$ = 0 and $\pi$ can be detected on other $k_z$ planes in the ARPES measurements. Thus, ARPES data are usually compared with the calculation results of the slab and the surface states.  The calculated surface states along $\bar{\Gamma}-\bar{K}$ are shown in Fig. \ref{3}(e). The observed surface band indicated by experimental data [Fig. \ref{3}(d)] corresponds well to the calculated surface states for the Sn-terminated (001) surface, as indicated in Fig. \ref{3}(e). Compared with the observations, the calculated bands around the nodal-line feature need to be relatively shifted up and down due to correlation effects of $d$-electron and interband coupling near $E\rm_F$ \cite{Liu2012,KFCA_Liu}.

In Fig. \ref{4}, we compare in detail the observed dispersions with the calculated surface states along the high-symmetry lines. Figures \ref{4}(a)-\ref{4}(c) show ARPES intensity plots along the $\Gamma$-$K$, $K$-$M$, and $\Gamma$-$M$ directions, respectively. The band crossings are clearly shown in the dashed rectangles in Figs. \ref{4}(a) and \ref{4}(b). As discussed above, the observed bands look more like the slab calculation with a Sn-terminated (001) surface. We thus show DFT-projected bulk bands and surface bands of the Sn-terminated (001) surface along the three corresponding directions in Figs. \ref{4}(d)-\ref{4}(f), respectively. The surface states and the Dirac nodes are indicated by the black arrows and the transparent dots, respectively. The calculation results are very consistent with the ARPES measurements including $k_z$ integration, except that the calculated holelike bands near $M$ need to be shifted down and the calculated nodal-line features near $K$ need to be shifted up for the reasons mentioned above. Combining the measured and calculated bands along $\Gamma$-$K$ [Figs. \ref{4}(a) and \ref{4}(d)], one can figure out the surface state (SS1) and the Dirac node (D1), which can be more clearly seen in the MDC plot in Fig. \ref{4}(g). D1 is located about 0.3 eV below $E\rm_F$, and SS1 corresponds to the marked surface states in Figs. \ref{3}(e) and \ref{4}(d). Because of the upward-shifting band, SS2 and D2 should be located above $E\rm_F$ and cannot be detected along $\Gamma$-$K$ and $K$-$M$. However, the consistency between experiment and calculation provides solid evidence of the existence of SS2 and D2 near $E\rm_F$. Compared with other nodal-line semimetals, the surface state (SS1) can be more clearly distinguished from the bulk states near $E\rm_F$. The results will supply valuable information for a further study of the Dirac nodal-line fermions.

\section{Conclusion}
By presenting an investigation of the low-energy electronic structures in both experiments and calculations, we provided evidence of topological nodal lines and surface states in centrosymmetric superconducting SnTaS$_2$. STS measurements indicate that a superconducting gap of about 0.35 meV at $T$ = 0.4 K is suppressed by a magnetic field of 0.013 T perpendicular to the sample surface. As typical characteristics in nodal-line semimetals, the surface states connecting the nodes make SnTaS$_2$ a good platform for exploring the topological nodal-line fermions. Our findings are also beneficial for studies on TSCs.

\section{ACKNOWLEDGEMENTS}
This work was supported by the National Key R\&D Program of the MOST of China (Grants No. 2016YFA0300204, No. 2016YFA0300200, and No. 2017YFA0303004), the National Science Foundation of China (NSFC; Grants No. 11704394, No. 11804176, No. 11888101, No. 11790310,  No. 11421404, No. 11674165, No. 11834006, and No. U2032208), and the Fundamental Research Funds for the Central Universities (Grant No. 020414380149).  D.W.S.  is  supported  by the Award  for  Outstanding Member in Youth Innovation Promotion Association CAS. The ARPES experiments were performed at the Dreamline beamline of SSRF and supported by the Ministry of Science and Technology of China (Grants No. 2016YFA0401002, and No. 2017YFA0403401) and the CAS Pioneer Hundred Talents Program. Part of this research used the 03U Beamline of the SSRF, which is supported by the ME2 Project (Grant No. 11227902) of NSFC.

\end{document}